\documentclass[twocolumn,preprintnumbers,amsmath,amssymb,superscriptaddress]{revtex4}
\usepackage[dvipdfmx]{graphicx}
\usepackage{color}
\usepackage{dcolumn}
\usepackage{bm}

\begin{document}

\title{Ce-doping and Reduction Annealing Effects on Electronic States in Pr$_{2-x}$Ce$_{x}$CuO$_4$\\Studied by Cu~{\it K}-edge X-ray Absorption Spectroscopy}


\author{Shun Asano}
\email{shun.asano@imr.tohoku.ac.jp}
\affiliation{Department of Physics, Tohoku University, Aoba, Sendai 980-8578, Japan}
\affiliation{Institute for Materials Research, Tohoku University, Katahira, Sendai 980-8577, Japan}
\author{Kenji Ishii}
\affiliation{Synchrotron Radiation Research Center, National Institutes for Quantum and Radiological Science and Technology, Hyogo 679-5148, Japan}
\author{Daiju Matsumura}
\author{Takuya Tsuji}
\affiliation{Materials Sciences Research Center, Japan Atomic Energy Agency, Hyogo 679-5148, Japan}
\author{Toshiaki Ina}
\affiliation{Japan Synchrotron Radiation Research Institute, Hyogo 679-5148, Japan}
\author{Kensuke M. Suzuki}
\author{Masaki Fujita}
\email{fujita@imr.tohoku.ac.jp}
\affiliation{Institute for Materials Research, Tohoku University, Katahira, Sendai 980-8577, Japan}

\begin{abstract}
We investigated Ce-substitution and reduction annealing effects on the electronic states at copper sites by Cu~{\it K}-edge x-ray absorption near-edge structure measurements in Pr$_{2-x}$Ce$_x$CuO$_{4+\alpha-\delta}$ (PCCO) with varying $x$ and $\delta$ (the amount of oxygen loss during annealing) values. Absorption near-edge spectra were modified by Ce-substitution and reduction annealing in a similar manner with increasing $x$ and $\delta$. Considering electron doping by Ce-substitution, this similarity indicates an increase of electron number at the copper sites due to annealing $n_{\rm AN}$. Thus, the total number of electrons is determined by the amount of Ce and oxygen ions. Furthermore, quantitative analyses of the spectra clarified that the number of Cu$^+$ sites, corresponding to the induced electron number by Ce-substitution $n_{\rm Ce}$ increases linearly with $x$ in the as-sintered PCCO ($\delta=0$), whereas $n_{\rm AN}$ is not exactly equal to twice of $\delta$, which is expected from charge neutrality. For each $x$-fixed sample, $n_{\rm AN}$ tends to exceed  2$\delta$ with increasing $\delta$, suggesting the emergence of two types of carrier due to annealing. 
\end{abstract}

\maketitle

\section{Introduction}
High transition temperature superconductivity occurs due to electron doping into $RE_{2-x}$Ce$_x$CuO$_4$ ($RE$ = Pr, Nd, Sm, Eu) with T'-type structure, where Cu ions have fourfold coplanar coordination. The undoped $RE_{2}$CuO$_4$ was reported to be an antiferromagnetic Mott insulator, as is the same ground state of La$_2$CuO$_4$, which is the parent compound of hole-doped superconducting La$_{2-x}$Sr$_x$CuO$_4$. As-sintered (AS) compounds of T'-$RE_{2-x}$Ce$_x$CuO$_4$ show insulating behavior, even in heavily electron doped region, and a post-annealing procedure in a reducing atmosphere is required for the emergence of superconductivity\cite{Takagi1989}. Several models have been proposed to explain the emergence of superconductivity due to annealing. Excess oxygen atoms may occupy the apical sites \cite{Radaelli1994} and/or Cu defects on the CuO$_2$ plane in the as-grown sample can be removed by annealing~\cite{Kang2007}. Although the local disorder is considered to be related to superconductivity, the microscopic mechanism of annealing-induced superconductivity and the variation in the electronic states due to annealing are not fully understood. 

Recently, a superconducting transition in T'-$RE_2$CuO$_4$ without cation substitution was reported to occur in annealed (AN) thin films~\cite{Matsumoto2009}. It is considered that the chemical disorder in the thin film can be adequately removed by annealing due to a large surface-to-volume ratio, whereas the removal of disorder in a homogeneous manner is difficult in a bulk sample. This means that the electronic states in T'-$RE_2$CuO$_4$ are sensitive to the chemical disorder, and undoped $RE_2$CuO$_4$ free from the disorder is not a Mott insulator. Subsequently, superconductivity in the parent T'-$RE_2$CuO$_4$ material was confirmed to occur in a low-temperature synthesized powder sample ~\cite{Takamatsu2012}. These results cast doubt on the fundamental recognition that parent compounds of the cuprate superconductors are antiferromagnetic Mott insulators. An important question then is regarding the influence of reduction annealing on the electronic states. 

Angle-resolved photoemission spectroscopy (ARPES) studies \cite{Song2017, Horio2015a, Wei2016} claimed that the electron number per Cu ion, which is estimated from the Fermi surface area, increases due to annealing and is larger than the concentration of Ce ($x$). This result suggests that the carrier concentration is influenced by non-stoichiometry of oxygen. However, the relationship between the concentration of oxygen vacancies and the carrier number varied by the annealing have not yet been investigated quantitatively. So far, the annealing effects were mostly studied in connection with the appearance of superconductivity, and no systematic measurement was reported for the undoped and lightly-doped bulk compounds, which do not show superconductivity even after annealing \cite{Oyanagi1990, Song2012}. In contrast to x-ray photoemission spectroscopy measurements, which are difficult to use for measuring the electronic states in insulating materials, x-ray absorption near-edge structure (XANES) measurements are suitable to evaluate electron-doping levels in compound types ranging from insulating to metallic \cite{Oyanagi1990, Kosugi1990, Asano2018}. 

In this work, we performed Cu {\it K}-edge XANES measurements on Pr$_{2-x}$Ce$_x$CuO$_{4+\alpha-\delta}$ (PCCO) with various $x$ and ${\delta}$ (the amount of oxygen loss). This is the first systematic XANES investigation on the evolution of the electronic states against oxygen contents at several $x$ values. 
It was found that variations in the absorption near edge spectra induced by reduction annealing are similar to the case of Ce-doping, indicating aspects of electron doping in the annealing effect. Furthermore, a detailed analysis of the spectra revealed that in as-sintered PCCO, the electron number increases linearly with $x$, whereas the number of additionally introduced electrons by annealing ${\it n_{\rm AN}}$ deviate from the simple relation ${\it n_{\rm AN}}$ = 2$\delta$ for each $x$ value. Thus, the annealing effects exhibit another phenomenon beyond electron-doping. 
 
	\begin{table}[b]
 	\begin{center}
    \caption{The amount of oxygen loss $\delta$ and reduction annealing conditions for Pr$_{2-x}$Ce$_x$CuO$_{4+\alpha-\delta}$.}
    \begin{tabular}{ccc} \hline
      $x$ & $\delta$ & Annealing conditions\\ \hline
      0 & 0.029(1) & 900$^\circ$C/12~hours \\ 
      0 & 0.040(1) & 940$^\circ$C/12~hours \\ 
      0.08 & 0.021(1) & 900$^\circ$C/12~hours \\ 
      0.08 & 0.031(1) & 940$^\circ$C/12~hours \\ 
      0.12 & 0.018(1) & 900$^\circ$C/12~hours \\ 
      0.12 & 0.023(1) & 940$^\circ$C/12~hours \\ 
      0.16 & 0.016(1) & 940F$^\circ$C/12~hours \\ \hline
    \end{tabular}
    \label{AN_condition_PCCO}
  	\end{center}
	\end{table}

\section{Sample preparation and XANES experiment}
Polycrystalline samples of PCCO were synthesized by a solid-state reaction method. 
Dried powders of Pr$_6$O$_{11}$, CeO$_2$ and CuO were mixed. 
The mixture was pressed into pellets and sintered at 1030${}^\circ\mathrm{C}$ in air with intermediate grinding. The oxygen-reduced samples were prepared by annealing the AS samples in flowing Ar gas at 900${}^\circ\mathrm{C}$ or 940${}^\circ\mathrm{C}$ for 12 hours. 
The phase purity of the samples was confirmed by laboratory-based x-ray powder diffraction measurements. The $\alpha$ and $\delta$ values represent the amount of excess oxygen in the AS samples and that of oxygen loss due to reduction annealing, respectively. 
Although we do not determine $\alpha$ for the present samples, $\alpha$ for the  samples previously prepared under the same conditions was evaluated to be 0.02--0.04 by iodometric titration. 
The value of $\delta$ was evaluated from the weight lost of the sample due to annealing, and it tends to be larger in samples annealed at higher temperature with fixed $x$. The value of $\delta$ and reduction annealing conditions are summarized in Table \ref{AN_condition_PCCO}. Among these samples, only PCCO with $x=0.16$ annealed at 940${}^\circ\mathrm{C}$, shows superconductivity with 18 K transition temperature ($T_{\rm c}$). XANES measurements were carried out at the BL01B1 and BL14B1 in SPring-8. Cu {\it K}-edge absorption spectra were measured in transmission mode using a Si(1 1 1) double-crystal monochromator. For each sample, we prepared a small pellet (7 mm in diameter and 0.5 mm in thickness) mixed with boron nitride so that the pellet was self-supported. All measurements were performed at 300~K using the pressed pellets.

	\begin{figure}[b]
	\begin{center}
	\includegraphics[width=80mm]{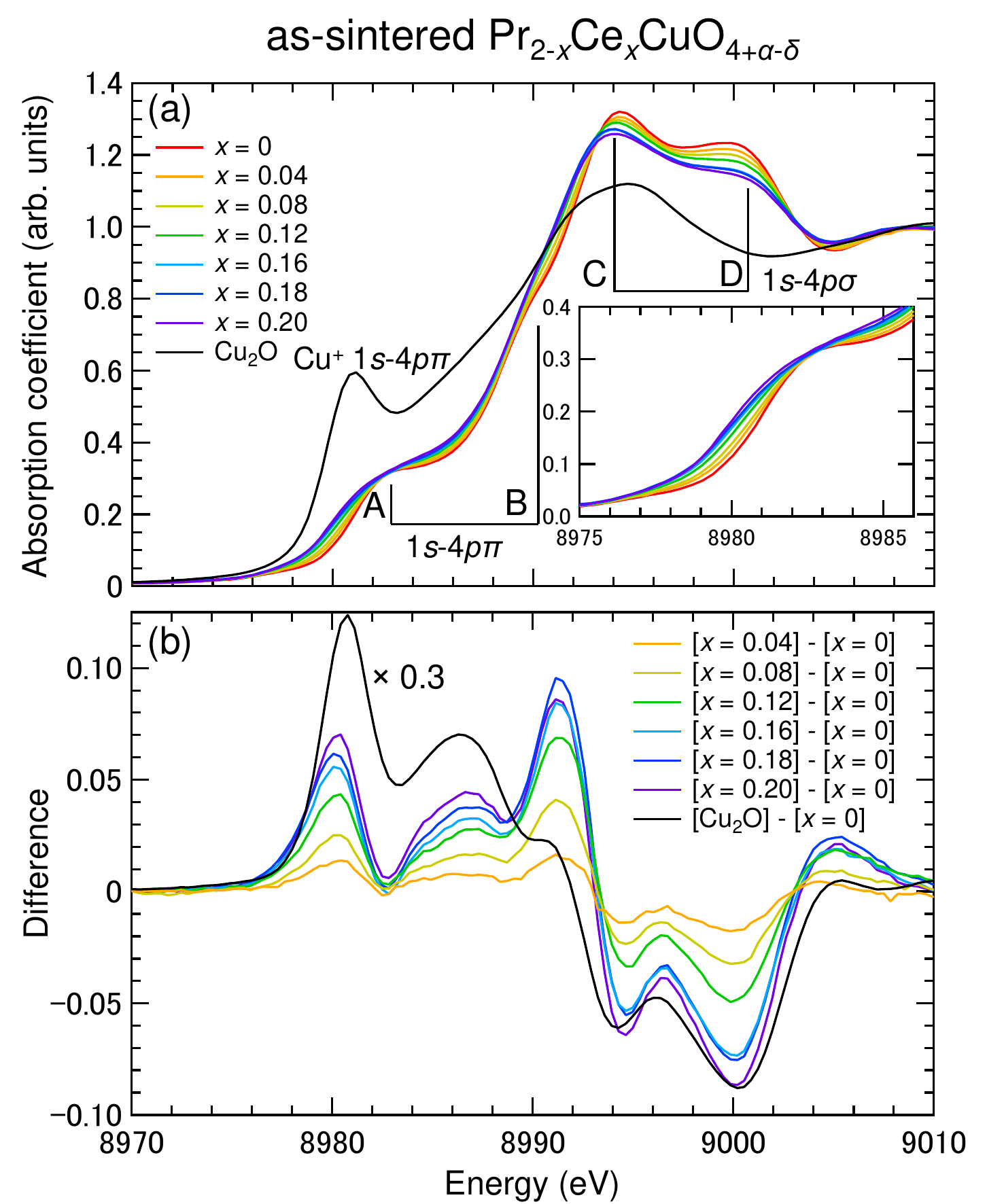}
	\caption{(Color online) (a) Cu~{\it K} absorption near-edge spectra for as-sintered (AS) Pr$_{2-x}$Ce$_x$CuO$_{4+\alpha-\delta}$ (PCCO) with $x$ = 0, 0.04, 0.08, 0.12, 0.16, 0.18, and 0.20. The spectrum of Cu$_2$O is also shown as a reference. The inset shows the spectra for energies between 8975~eV and 8986~eV. (b) The spectrum for AS~PCCO $x = 0$ is subtracted from spectra for AS~PCCO with $x \geq 0.04$ and Cu$_2$O. The difference spectrum for Cu${_2}$O shown after being multiplied by 0.3.}
	\label{PCCO_AS}
	\end{center}
	\end{figure}

	\begin{figure*}[htb]
	\begin{center}
	\includegraphics[width=180mm]{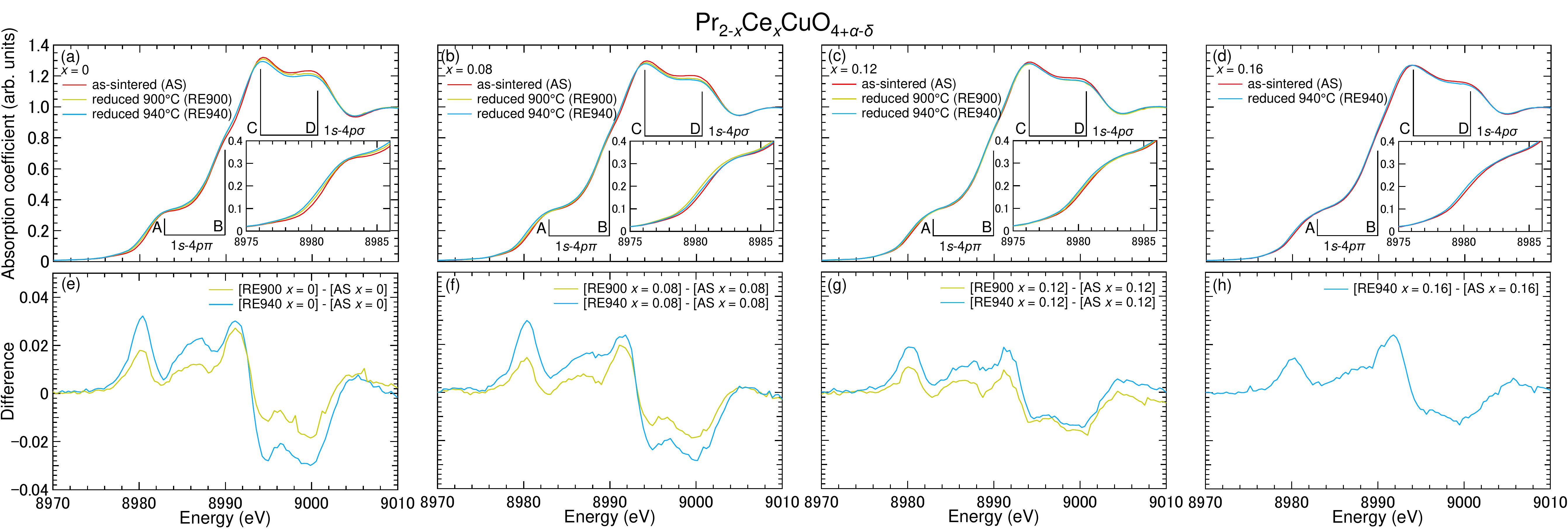}
	\caption{(Color online) (a)--(d) Cu~{\it K} absorption near-edge spectra for as-sintered (AS), reduced at 900${}^\circ\mathrm{C}$ (RE900), and reduced at 940${}^\circ\mathrm{C}$ (RE940) Pr$_{2-x}$Ce$_x$CuO$_{4+\alpha-\delta}$ (PCCO) with $x$ = 0, 0.08, 0.12, and 0.16. Inset shows the spectra for energies between 8975~eV and 8986~eV. (e)--(h) Difference spectra for RE900 and RE940~PCCO.}
	\label{PCCO_AN_all_v2}
	\end{center}
	\end{figure*}

\section{Results}
Figure \ref{PCCO_AS}(a) shows Cu {\it K} absorption near-edge spectra for AS PCCO with $x$ = 0--0.20. The intensity of the spectra for AS PCCO is normalized by the values in the high-energy region, which is insensitive to the electronic states. The characteristic features in the spectra are assigned in accordance with the previous studies as follows\cite{Oyanagi1990, Kosugi1990, Asano2018}. The shoulders at 8983 eV (labeled A) and 8991 eV (labeled B) correspond to dipole transitions from 1{\it s} to 4{\it p}$\pi$ of Cu$^{2+}$, and the peaks at 8994 eV (labeled C) and 9000 eV (labeled D) correspond to dipole transitions from 1{\it s} to 4{\it p}$\sigma$ of Cu$^{2+}$. The final states for A and C are well-screened states, and those for B and D are poorly-screened states. 
The spectrum for Cu$_2$O is also plotted as a reference of Cu$^+$ configuration.
When a core hole is created at the Cu$^+$ site, core hole potential at the site is screened by the same number of electrons as the well-screened state.
While cost of the charge transfer energy is necessary for the well-screened state, it is not for the state at the Cu$^+$ site.
Then, peak of the Cu$^+$ state is located at lower in energy than the well-screened state \cite{Liang1995, Tsutsui2003} and we ascribe the peak at 8981~eV to the energy of 1{\it s}--4{\it p}$\pi$ transition of Cu$^+$ in Cu$_2$O.
		
As seen in Fig. \ref{PCCO_AS}(a), the spectra around the near-edge region vary systematically upon Ce-substitution. Intensity around an energy corresponding to 1{\it s}--4{\it p}$\pi$ transitions gradually increase, while that for 1{\it s}--4{\it p}$\sigma$ transitions decrease with increasing Ce concentration. Figure \ref{PCCO_AS}(b) shows the difference spectra, which are obtained by subtracting the normalized spectrum of AS PCCO with $x$ = 0 from that of AS PCCO with $x$ = 0.04--0.18. The difference spectra of AS PCCO show a peak at an energy similar to the 1{\it s}-4{\it p}$\pi$ transition in Cu$^+$ in Cu$_2$O, indicating the formation of Cu$^+$ sites by Ce-substitution. The intensity enhancement indicates increased electron density at the copper sites, which is consistent with the result from a Cu $L$-edge x-ray absorption spectroscopy study reporting the reduction of unoccupied states in Cu 3$d_{{x^2}-{y^2}}$ orbitals \cite{Fillipse1990}. 

We next investigated the effects of reduction annealing on the near-edge structure of Cu {\it K}-edge absorption spectra. In Fig. \ref{PCCO_AN_all_v2}(a), the spectra from PCCO with $x$ = 0 reduced at 900${}^\circ\mathrm{C}$ (RE900) and at 940${}^\circ\mathrm{C}$ (RE940) are shown together with that for AS PCCO with $x$ = 0. The absorption spectra vary with reduction annealing temperature. The intensity at A and B (C and D) corresponding to 1{\it s}-4{\it p}$\pi$ (1{\it s}-4{\it p}$\sigma$) transitions is stronger (weaker) with larger $\delta$. The difference spectra between the AN and AS samples are shown in Fig. \ref{PCCO_AN_all_v2}(e). The structure of the difference spectra is quite similar to that induced by Ce-substitution (see Fig. \ref{PCCO_AS}(b)). It is clear that the reduction annealing increases electrons at the Cu sites. As seen in Figs. \ref{PCCO_AN_all_v2} (b)--(d) and (f)--(g), PCCO with $x=0.08$, 0.12, and 0.16 show the same trend against $\delta$, suggesting a common aspect of electron doping in the reduction annealing for PCCO. 

\begin{figure}[t]
	\begin{center}
	\includegraphics[width=85mm]{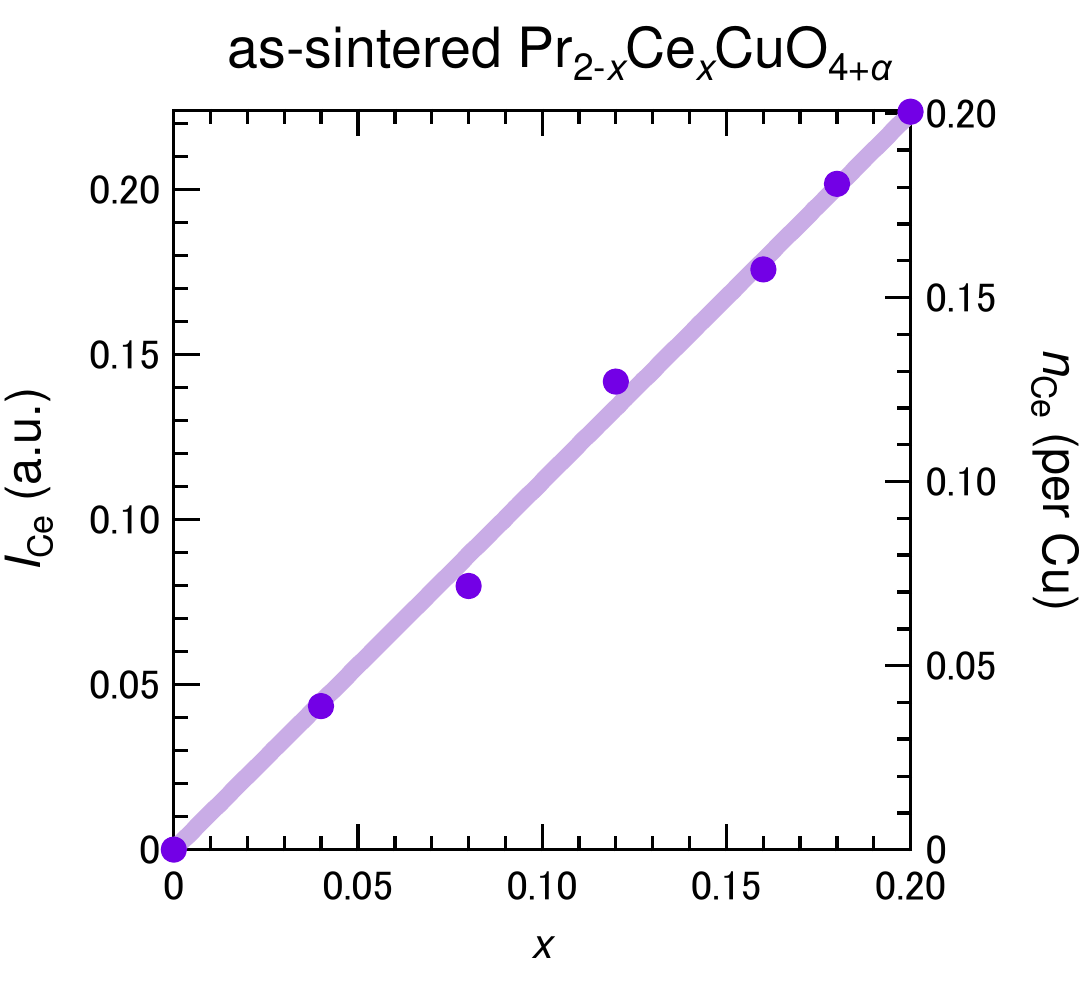}
	\caption{(Color online) Increased intensity of the Cu$^+$~1{\it s}-4{\it p}$\pi$ dipole transitions $I_{\rm Ce}$ as a function of Ce content $x$ for as-sintered (AS) Pr$_{2-x}$Ce$_x$CuO$_{4+\alpha-\delta}$ (PCCO). $I_{\rm Ce}$ is estimated by integrating the difference spectra between 8976 eV and 8993 eV. The right vertical axis represents the electron number per Cu atom $n_{\rm Ce}$. The solid line is a fit to a linear function with fixed $I_{\rm Ce}=0$ at $x=0$.}
	\label{Int_vs_Ce}
	\end{center}
	\end{figure}

A relative variation of the electron number is evaluated by an integration of the absorption spectra between 8976~eV and 8983~eV, which corresponds to the intensity of Cu$^+$ 1{\it s}-4{\it p}$\pi$ dipole transitions, and this is expected to be proportional to the number of Cu$^+$ sites. The $x$-dependence of $I_{\rm Ce}$, which is defined as the increased amount of the intensity integrated between 8976~eV and 8983~eV by Ce-doping into AS Pr$_2$CuO$_4$, is shown in Fig. \ref{Int_vs_Ce} for AS PCCO. As one of the results in this study, we found that $I_{\rm Ce}$ increases linearly with increasing Ce-substitution. Considering the previous optical conductivity studies, which show a clear energy gap (charge transfer gap) in the optical spectra for the AS Pr$_2$CuO$_4$ and Nd$_2$CuO$_4$, indicating there are no or negligible carriers in the AS samples\cite{Arima1993, Onose2004}. Furthermore, neutron scattering experiments reported that the evolution of 
magnetism by Ce-substitution in the AS sample can be well understood by a spin dilution model~\cite{Keimer1992,Mang2004a}, indicating an identical relationship between the proportion of Cu$^+$ sites induced by Ce-substitution and the Ce content, $x$. 
Based on these experimental results, we determined the conversion factor between ${n_{\rm Ce}}$ and $I_{\rm Ce}$ from a linear fit for the $x$-$I_{\rm Ce}$ relation of AS PCCO with fixed $I_{\rm Ce}=0$ at $x=0$ under the assumption that ${n_{\rm Ce}}$ is equal to $x$. Here, ${n_{\rm Ce}}$ is the increased number of electrons per Cu atom by Ce-doping into AS Pr$_2$CuO$_4$. The right vertical axis in Fig. \ref{Int_vs_Ce} represents ${n_{\rm Ce}}$, and the relation between ${n_{\rm Ce}}$ and $I_{\rm Ce}$ can be written as ${n_{\rm Ce}}=I_{\rm Ce}/1.116$. 

\begin{figure}[t]
	\begin{center}
	\includegraphics[width=85mm]{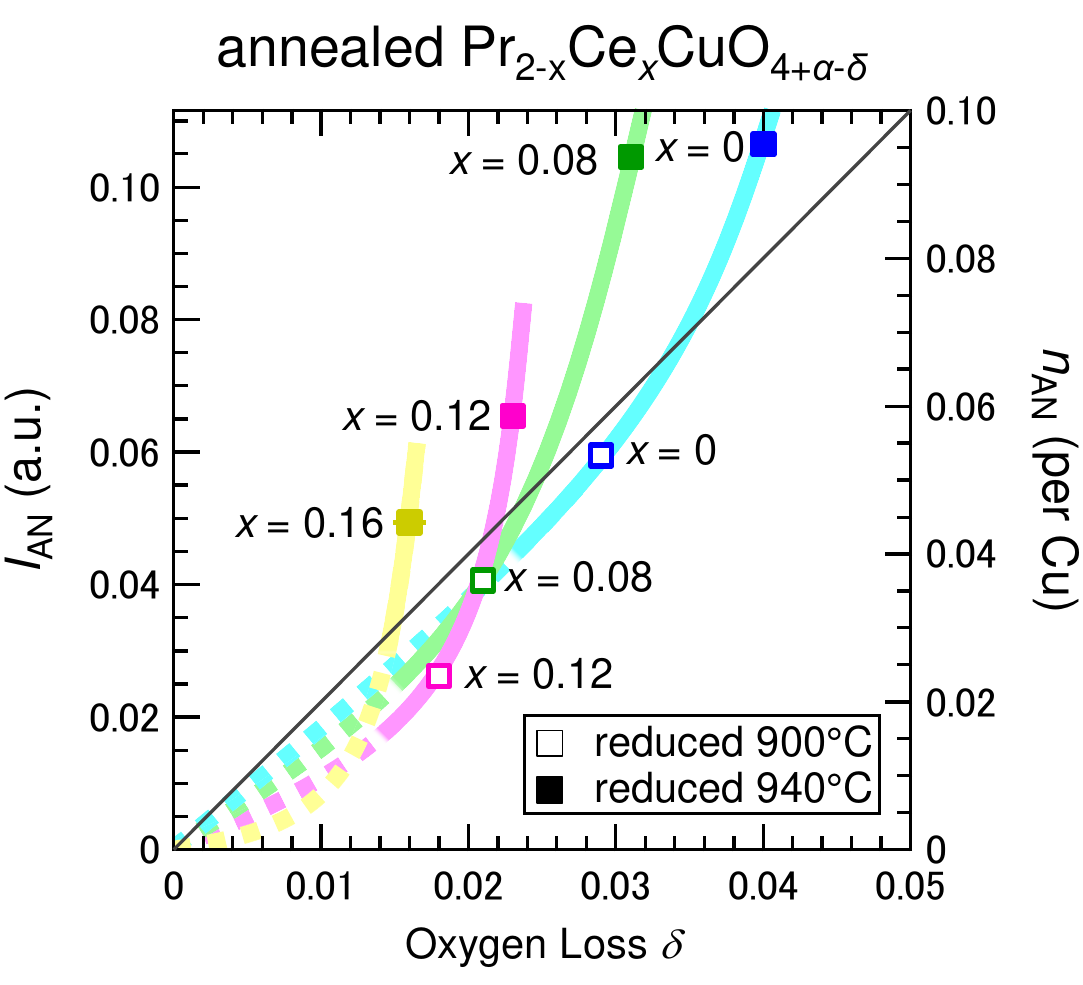}
	\caption{(Color online) The increased amount of the intensity of Cu$^+$ 1{\it s}-4{\it p}$\pi$ dipole transition, ${\it I_{\rm AN}}$, (the left vertical axis) and that of the electron number per Cu atom, ${\it n_{\rm AN}}$, (the right vertical axis) through the annealing as a function of the oxygen loss $\delta$ for RE900 and RE940 Pr$_{2-x}$Ce$_x$CuO$_{4+\alpha-\delta}$ (PCCO). The gray solid line represents a relation of ${\it n_{\rm AN}} = 2\delta$. The solid and dashed curved lines are the guide to the eye for RE900 and RE940 PCCO with $x$ = 0 (cyan), 0.08 (green), 0,12 (magenta) and 0.16 (yellow). 
}
	\label{Int_vs_Delta_v6}
	\end{center}
	\end{figure}

Figure \ref{Int_vs_Delta_v6} shows the increased intensity for the Cu$^+$ 1{\it s}-4{\it p}$\pi$ dipole transitions due to annealing in the sample with fixed $x$, ${\it I_{\rm AN}}$ (the left vertical axis). The figure also shows the electron number per Cu atom introduced by annealing ${\it n_{\rm AN}}$ (right vertical axis) as a function of the oxygen loss, $\delta$, for RE900 and RE940 PCCO. The ${\it n_{\rm AN}}$ value is evaluated from the above relationship between ${n_{\rm Ce}}$ and $I_{\rm Ce}$. The ${\it I_{\rm AN}}$ (${\it n_{\rm AN}}$) tends to increase as $\delta$ increases. The gray solid line in the figure represents a relationship of ${\it n_{\rm AN}} = 2\delta$, which is expected from charge neutrality of the sample with the assumption that all electron carriers are doped into the CuO$_2$ plane. The present data roughly obeys this relation, meaning that the introduction of electrons through annealing comes from the removal of oxygen. However, when we look at the $\delta$-dependence of the ${\it I_{\rm AN}}$ (${\it n_{\rm AN}}$) for each $x$, a characteristic tendency becomes visible. In Fig. \ref{Int_vs_Delta_v6}, solid curves are drawn to guide the eye and illustrate the relationship between $\delta$ and ${\it n_{\rm AN}}$ for samples with fixed $x$. Here, ${\it n_{\rm AN}}$ should be zero at $\delta$ = 0. In a sample with larger $x$, the variation of ${\it n_{\rm AN}}$ is more sensitive to $\delta$; ${\it n_{\rm AN}}$ for $x$ = 0 approximately corresponds to 2$\delta$, while that for $x$ = 0.12 increases with increasing $\delta$, including a marked change around $\delta$ = 0.02. The slope coefficient seems to be larger for the sample with larger $x$. These results are difficult to understand within the simple spin dilution model, as is not the case for Ce-substitution. 

\section{Discussion}
In the present study, we clarify the similarities and differences between Ce-substitution and annealing effects on the electronic states at the copper sites by Cu~{\it K}-edge XANES measurements. Both Ce-substitution and annealing effects exhibit an aspect of electron-doping. Although the Ce-substitution effects for the AS sample are understood under the spin dilution model, the intensity enhancement due to annealing ${\it I_{\rm AN}}$ is rather complicated. Here, we discuss how annealing may affect the electronic states. 

We first mention that Cu~{\it K}-edge XANES measurements are used to collect information on the doped electrons through the creation of Cu$^+$ sites. No or less information about the holes, which would be predominantly doped at oxygen sites, is detected. 
Indeed, the effects of hole-doping on the Cu {\it K} absorption near-edge spectra in La$_{2-x}$Sr$_x$CuO$_4$ is reported to be much smaller than the electron-doping effects in Nd$_{2-x}$Ce$_x$CuO$_4$~\cite{Kosugi1990}. Considering this result and charge neutrality, the present results located in the ${\it n_{\rm AN}}$ $>$ 2$\delta$ region (the region above the straight line) in Fig. 4 indicates the existence of holes which compensate a number of induced electrons in excess of ${\it n_{\rm AN}}$ = 2$\delta$ in the sample. Thus, the deviation from ${\it n_{\rm AN}}$ = 2$\delta$ toward the upper side suggests an additional aspect in reduction annealing beyond the simple picture of electron-doping by Ce-substitution. On the other hand, in the ${\it n_{\rm AN}}$ $<$ 2$\delta$ region (the region below the straight line), some of the electrons induced by annealing are not introduced into the Cu site. 
Thus, an upturn in the $\delta$-${\it n_{\rm AN}}$ relation indicates that as $\delta$ increases, the carrier characteristics change from the region where electrons do not effectively enter the CuO$_2$ plane to the region where two kinds of carriers are considered to exist. The variation tends to take place at smaller $\delta$ values in a sample with larger $x$, suggesting the existence of a critical electron number required to induce the two kinds of carriers. The two types of carrier in T'-$RE_{2-x}$Ce$_x$CuO$_4$ was suggested from the Hall coefficient and Hall resistivity measurements on thin films\cite{Dagan2004,Gauthier2007}, 
although the amount of removed oxygen did not determined for these samples.
From the present systematic study, we showed experimental evidence that reduction annealing plays an essential role in the emergence of hole carriers, and that holes can appear with smaller $\delta$ for a sample with larger $x$. 
We speculate that such a change originates from a band structure transformation via heavy electron doping. 
Thus, precise and systematic ARPES measurements on high quality samples, in which the amount of oxygen loss is well-controlled, are indispensable to determine the existence and origin of two types of carriers. 

We now focus attention on the electronic states in the AS and AN samples with large and comparable electron numbers. Compared to AS PCCO with $x=0.20$ and RE940-PCCO with $x=0.16$, the integrated absorption spectra between 8976 eV and 8983 eV, i.e., the total value of $I_{\rm Ce}$+${\it I_{\rm AN}}$, is almost equivalent. 
Thus, the total electron number in these compounds is comparable, although the former sample is an insulator and the latter one is a superconductor with $T_{\rm c}$ of 18 K.
The difference between the ground state in the SC(AN) and non-SC(AS) samples is mostly discussed in connection with the presence/absence of chemical disorder. The random potential in the CuO$_2$ plane suppress superconductivity. 
As for other differences, as mentioned above, the AN sample with larger $\delta$ would contain both electron and hole carriers, whereas the results for the AS sample is explained by the spin dilution model, which considers only electrons with $n_{\rm Ce}$ = $x$. Therefore, the electronic states relating to the hole carries in these two compounds might be different. 

From this point of view, distinct ground states in between the SC thin film and non-SC bulk samples of parent T'-$RE_2$CuO$_4$ are possibly related with the existence or amount of hole carriers, although a considerable number of electrons can be doped in both film\cite{Horio2017, Horio2018a} and bulk samples by annealing. In the present PCCO with $x$ = 0 and $\delta$ = 0.040, in which the total electron number of $n_{\rm Ce}$ + $n_{\rm AN}$ is 0.094, the deviation of ${\it n_{\rm AN}}$ from the relation ${\it n_{\rm AN}}$ = 2$\delta$ yields the hole number $n_{\rm h}$ of 0.014. 
In order to clarify the necessary conditions for the emergence of superconductivity and for the existence of multiple carriers 
in T'-$RE_{2-x}$Ce$_x$CuO$_4$, the electron and hole numbers for both SC thin film and a bulk sample with various $x$ and $\delta$ values must be determined.

\section{Summary}
	In summary, Ce-substitution and reduction annealing effects on the electronic states at copper sites were investigated by Cu {\it K}-edge x-ray absorption near-edge structure measurements for T'-type Pr$_{2-x}$Ce$_x$CuO$_{4+\alpha-\delta}$. For the as-sintered PCCO, the absorption near-edge spectra systematically changed with Cu$^+$ content, and the electron number increased linearly with Ce-substitution. The spectra were changed by reduction annealing, similar to the case for Ce-substitution over the entire $x$ range in the present study. This result indicates the aspect of electron doping in reduction annealing. However, the electron number increased due to annealing $n_{\rm AN}$ does not exactly follow the relation of $n_{\rm AN}$ = 2$\delta$, where $\delta$ is the amount of oxygen lost during annealing. Considering the charge neutrality, the upper deviation of $n_{\rm AN}$ from $n_{\rm AN}$ = 2$\delta$ suggests the emergence of two types of carrier due to annealing. 

\section*{Acknowledgements}
We thank T. Adachi for the fruitful discussion and Y. Kimura for the support in the analysis. The synchrotron radiation experiments were performed at the BL01B1 of SPring-8 with the approval of the Japan Synchrotron Radiation Research Institute (JASRI) (Proposal No. 2016A1603 and No. 2017B3611).  M.F. is supported by Grant-in-Aid for Scientific Research (A) (16H02125) and K.I. is supported by Grant-in-Aid for Scientific Research (B) (16H04004).
	
\bibliography{XAFS_PCCO.bib}

\end{document}